\documentstyle[twocolumn,epsf,floats,prb,aps]{revtex}
\include{psfig}
\newcommand{\figwidth}{3.375in}
\begin{document}

\twocolumn[\hsize\textwidth\columnwidth\hsize\csname @twocolumnfalse\endcsname

\title{Quantum Films Adsorbed on Graphite:  Third and Fourth Helium Layers}
\author{Marlon Pierce and Efstratios Manousakis}
\address{
Department of Physics and Center for Materials Research and Technology,
Florida State University, Tallahassee, FL 32306-4350
}
\date{\today}
\maketitle
\begin{abstract}
\noindent
Using a path-integral Monte Carlo method for simulating superfluid 
quantum films, we investigate helium layers adsorbed on a substrate 
consisting of graphite plus two solid helium layers.  The solid helium
layers are modeled first as inert, with paths frozen at equilibrated
positions, and then as active, with second layer atoms included in the
Monte Carlo updating.  In both cases, we observe
the formation of as many as three well defined additional layers above the first two,
and determine the layer promotion density by calculating the 
density profile and through a calculation of the
chemical potential.  For liquid layers
adsorbed onto the inert solids, we
find self-bound liquid phases in both the third and fourth layers and
determine the equilibrium density.  In the third layer at coverages
below equilibrium, we find liquid droplets and a metastable uniform liquid
phase, and determine the spinodal point that separates these regions.  
The above phases and their coverage ranges are in good agreement with
several experiments.
The superfluid density as a function of coverage is also calculated, and it
is observed to change only weakly around the promotion density.  For 
coverages above the beginning of fourth layer promotion, we observe
continued increase in the third layer density.  We note that the third
layer density increase is perhaps enough to cause solidification in
this layer, which would explain heat capacity peaks observed experimentally
for fourth layer coverages and would provide a simple explanation for the
plateaus seen in the superfluid coverage.
For helium 
adsorbed on an active second layer, we observe that a self-bound
liquid phase occurs in the third layer, and we determine
the equilibrium density and spinodal point,
which remain in agreement with experiment.  We find that promotion
to both the third and fourth layers is signaled by a change in the
density dependence of the chemical potential.
We further observe the increase
in the second layer density with increasing total coverage.  The coverage 
dependence of the superfluid density is calculated, and a 
pronounced drop is
seen at high third layer coverages, as has also been observed
experimentally.  
\end{abstract}
] 
\section{Introduction}
\label{sec:intro_layr3}
Helium films adsorbed on graphite exhibit a number of phases and 
have proven to be a rich source for both experimental and theoretical studies
of two-dimensional phenomena.  The graphite substrate is ordered
on atomic length scales and offers a potential well for helium that
is relatively strong for physical adsorption but is
short ranged perpendicular to the substrate.  
As a result, a number of distinct, atomically thin layers occur, each with its
own phase diagram.  Near the graphite surface, the layers tend
to solidify, 
with both commensurate and incommensurate solids occuring in 
the first two layers.  In both layers, this solidification 
occurs before promotion to the next layer.  The
second layer exhibits a coverage region with superfluidity as 
well, \cite{reppy93,reppy96,pierce98,pierce99a}
while the first layer apparently favors the formation of solid clusters over
liquid droplets at low densities, \cite{ecke85,pierce99b,pierce00a} a
lthough there is 
debate on this issue.\cite{greywall93,bruch93}  A general review of
physically adsorbed films such as helium on graphite
can be found in the book of 
Bruch, Cole, and Zaremba.\cite{bruch97}
In this paper, we will focus on the liquid third and fourth
layers.  

Detailed information on the film structure at low temperatures for the third 
and higher layers has come from a number of experiments, including
heat capacity,\cite{greywall91,greywall93}
torsional oscillator,\cite{reppy93,reppy96,saunders98} 
and third sound measurements.\cite{zimmerli92}  
Ranges in which the heat capacity depends linearly on the coverage
suggest that gas-liquid coexistence regions exist
in the third and fourth layers. 
The transition from liquid droplets to a uniform liquid phase in each
layer is  
signaled by a peak in the isothermal compressibility.\cite{zimmerli92}
Unlike the first and
second layers, these higher layers do not 
solidify before layer promotion, since
torsional oscillator measurements detect superfluidity for all 
coverages beginning at intermediate third layer densities.  This apparently
rules out earlier suggestions that the third layer may 
solidify,\cite{greywall93,zimmerli92}  although it still may be possible for
solidificatino to take place under compression of higher layers.


Perhaps the most unusual feature of the higher layers is the step-like behavior
of superfluidity with increasing 
density.\cite{reppy93,reppy96}  The superfluid coverage
in layered films will not grow continuously because of the layering
transitions.  Plateaus in the superfluid density 
immediately after layer promotion are expected, and have been observed,
when the particles are in 
the droplet region.  These plateaus occur because the
droplets in the new layer lack the connectivity to exhibit superflow
across the entire surface.\cite{dash78a}
The interesting observation is that
the plateaus actually begin {\em before} layer promotion for the third 
through sixth layers.
At 500 mK, superfluidity even exhibits a decrease with increasing coverage 
near the promotion to the fourth layer.
These effects have been discussed in the context of the Bose-Hubbard
model.\cite{zimanyi94}  The suggestion is that the plateaus are produced
by the increased localization of particles in the dense liquid. 

Theoretical tools applied in the study of quantum films 
on realistically treated substrates include
the hypernetted chain Euler-Lagrange (HNC-EL) 
method,\cite{clements93a,clements93b,saslow96,campbell97}
density functional theory,\cite{treiner91,cheng92,cheng93} and 
quantum Monte Carlo.
\cite{abraham87,abraham90,wagner94a,wagner94b,bruch93,pierce98,pierce99a}  
The HNC-EL approach has been used extensively
in studies of the third and higher layers of helium films on graphite.
This approach determines stable coverages of the helium layers.  
Because the theory requires a uniform film, calculations
cannot be made for all coverages.  Breakdowns in the theory are
interpreted as occuring at coverages where
the film is unstable to the formation of droplet patches or 
to layer promotion.  The theory predicts at least 
three liquid layers will form
on top of the solid first and second layers, and yields a maximum coverage
value before promotion of 0.065 atom/$\AA^2$ for each layer, in good 
agreement with, but
somewhat below, the experimental value of 0.076.\cite{greywall93,zimmerli92}

In this paper we present results for the third and fourth helium layers using
the path-integral Monte Carlo (PIMC) method.  
Our simulation is able to take into account the
effect of the second layer's corrugations and zero-point motion on the 
third layer.  
We are also able to allow for the possibility of promotion and demotion
of particles between the second, third, and fourth layers,
and these effects are
observed.  Finally, our simulation method can be applied to the
entire range of possible phases in a layer, from liquid droplets to
full solidification.  Thus we are able to probe both the low
density and high density phases of a layer.

-----------------------------------------------------------------------

\section{Simulation method}
\label{sec:method}

Our simulation is performed using a path-integral Monte Carlo method
that includes particle permutations and substrate effects.  The general
method for bulk simulation has been reviewed elsewhere, \cite{ceprev} and
our modifications for simulating layered systems are given in a previous
publication.\cite{pierce99a}  Some of the present calculations have
required minor changes to this method, so we will briefly outline the
procedure now in order to explain the modifications.

\subsection{Overview of procedure}
The partition function $Z$ for a system of $N$ bosons at 
the inverse temperature $\beta$ can be expanded as a path-integral 
by inserting $M$ intermediate configurations:
\begin{eqnarray}
Z \ = & & \frac{1}{N!} 
	\sum_P \int...\int d^3 R_1 ... d^3 R_M d^3 R \nonumber \\
  & & \times \rho ({\bf R}_1,{\bf R}_2; \tau) \rho({\bf R}_2, {\bf R}_3; \tau) 
  	\ldots \rho ({\bf R}_M,P {\bf R}_1; \tau),
\label{eq:pathint}
\end{eqnarray}
where $\rho$ is the density matrix, ${\bf R}_i$ is a configuration of 
$N$ particles, and $\tau=\beta/M$.
The sum over $P$ is the sum of all possible permutations of particle labels.
Both the configurations and the permutations are sampled by our PIMC method.
The number of intermediate configurations used, $M$, is referred to as
the number of inverse-temperature slices.

To implement this method, two important ingredients are needed.  First,
a starting approximation for the density matrix at $\tau$ is required.  The
simplest of these 
is the so-called semiclassical approximation.  The drawback to this
approach is that very high starting temperatures are needed to obtain
an accurate approximation.  Methods for improving the starting approximation
can lower the starting temperature 
to $\tau^{-1}=40 K$.\cite{ceprev}  

The second important feature of the method is multilevel sampling.  The
path-integral, Eq.\ (\ref{eq:pathint}), can be thought of as a system 
of $N$ ring
polymers, each with $M$ beads.  Particle permutations correspond to
splicing the ring polymers together.  In multilevel sampling, we update
particle positions over a section of the polymer chain.  This allows us
to implement permutations by spreading them over several 
inverse-temperature slices.  The number of slices to be updated is
$2^l$, where $l$ is the level of the move.  The value of $l$ must be chosen
to balance the acceptance of particle moves and particle permutations.
Increasing the value of $l$ increases the rate of accepting permutations
but decreases the acceptance of new particle positions, while decreasing
$l$ has the opposite effect.
Typically, we take $l=3$ since this gives the optimum balance between 
the acceptance of permutations and new particle positions.

\subsection{Third and higher layers}
In our calculations for the third and higher layers, we use a 
simulation cell that is 
designed to accommodate the first layer helium solid.  Periodic
boundary conditions are applied in the plane of the substrate.
We approximate the
effects of the solid first layer on the second and higher layer
atoms by placing frozen atoms at triangular lattice
sites on the substrate.  These frozen atoms are located 2.8$\AA$ above
the graphite surface, the height of the first layer as indicated
by neutron scattering.\cite{carneiro81}
The second layer also solidifies before promotion
to the third layer occurs.  In a previous simulation,\cite{pierce99a} we
determined that the highest second layer coverage before layer promotion
was 0.2117 atom/$\AA^2$.  This simulation was performed with 20 
second layer atoms above 30 frozen first layer atoms
in a cell of dimensions $15.075\AA \times 15.67 \AA$.  The majority
of our simulation results for the third and higher layers were obtained
using a simulation cell of this size.

In order to investigate the effects of an
inert versus active second layer, we have performed two sets of calculations.
In the first, we equilibrate the second layer at  
0.2117 atom/$\AA^2$.  The positions of these
atoms are then frozen and
are no longer included in the sampling.  Additional atoms are then placed 
above this inert substrate and have their positions and permutations
sampled.  We refer to these additional atoms as being ``active''.
The bisection 
level used in this simulation was $l=3$.  All errors that we report
for these calculations are statistical errors arising from
the Monte Carlo simulation of the active atoms above a
particular frozen
second layer.  There will be additional systematic errors that arise
from our particular choice for the frozen second layer configuration.

In the second set of calculations, both the second and third layer
particles are included in the sampling.  
There is a potential problem with doing Monte Carlo calculations on such
a system.
The third layer is liquid and permutations will occur at low
temperatures.  This favors using
$l=3$ for this layer.  The second layer, on the other hand, is solid and
furthermore increases in coverage as the overall coverage is increased.  
Using $l=3$ produces a low acceptance
rate for particle moves in the compressed second layer.  Sampling
efficiency can be improved by using $l=2$ for this layer.  Thus a single
value of $l$ for the entire system is not optimum.

To give each layer the best value of $l$, we have partitioned
atoms into ``second layer'' and ``third layer'' atoms.  ``Second layer''
atoms have their positions initially taken from an equilibrated second
layer solid.  ``Third layer'' atoms are started from an initial configuration
of atoms placed at various heights above the second layer.  
Promotion and demotion between the layers are allowed, but the layer label of
the atoms does not change.  ``Third layer'' atoms
are sampled with $l_{3rd}=3$, while ``second layer'' atoms are 
sampled at $l_{2nd}=2$.  
Permutations are allowed between atoms with the same layer label, but we
do not allow atoms with different layer labels to permute.    
This limitation is not a problem for atoms demoted to the second layer 
after being
initially placed on the third layer, since
exchanges are uncommon in the solid second layer. 
The promotion of a particle from 
the second layer to the third occasionally occurs after a particularly long
simulation.  At most, only one particle is promoted, so this will
have little effect on the permutations in the third layer in all but the
lowest coverages.
This approach gives a higher acceptance rate for
second layer particle moves, while allowing permutations to occur reasonably
often in the third layer.  We have found that this partitioning of $l$ 
lowers the energy by an amount ranging from
0.0 K to 0.1 K per atom relative to energy
calculations with a single value $l=3$ for the entire system.  The smallest
energy shift occurred at the lowest coverage tested, 0.2286 atom/$\AA^2$.
At this coverage the two values were within error bars.
The greatest energy shift occurred near the equilibrium coverage
of the third layer liquid.  This was the highest coverage we tested.  We 
also compared the energy values at selected coverages to calculations 
performed with a
single value $l=2$ for the entire system and found them to agree,
but third layer superfluidity was suppressed, as expected.

\section{RESULTS FOR THE THIRD AND FOURTH LAYERS}
\label{sec:layr3}

Before presenting our simulation results, 
we wish to clarify the convention that we use to report 
adsorbed helium coverages.
Normally, the values we give for the density
are for the total adsorbed helium
(first two solid layers plus any additional coverages for the higher layers).
Relative coverages within a layer are prefaced with a reference to that layer.
For example, the coverage 0.2583 atom/$\AA^2$ using our standard
simulation cell corresponds to a third layer coverage of 0.0466 atom/$\AA^2$
plus two solid layers with a combined coverage of 0.2117 atom/$\AA^2$.  

\subsection{Results with the inert second layer}
\subsubsection{Layer promotion and demotion}
\begin{figure}[htp]
\epsfxsize=\figwidth\centerline{\epsffile{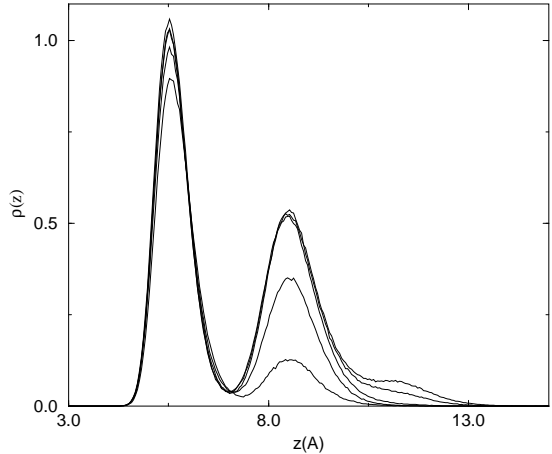}}
\caption{Density profiles at 400 mK for the second, third, and 
fourth layers as a function of height above the graphite substrate.  
The leftmost peak is for the frozen, equilibrated second layer.  The
coverages shown begin at 0.2329 atom/$\AA^2$ and increase in increments 
of 0.085, up to
0.3345.  The density profile for 0.3811 atom/$\AA^2$ is also shown.
The profiles are normalized so that integration gives the number of 
atoms.  
}
\label{fig:zdist}
\end{figure}

In order to study multiple layers of the helium film, it is first necessary
to establish that our simulation does, in fact, produce distinct layers
with increasing coverage.  This is illustrated in 
Fig.\ \ref{fig:zdist}, which shows the growth of the 
density profile perpendicular
to the substrate for the third and higher layers with increasing density. 
The peaks associated with the third and fourth 
layers can be clearly observed, as can the beginning of the fifth layer
peak at the highest simulated coverage.  
Also as can be seen in 
the figure, the continued growth of 
the third layer peak for all coverages indicates compression of this layer
even after atoms are promoted to the higher layers.

\begin{figure}[htp]
\epsfxsize=\figwidth\centerline{\epsffile{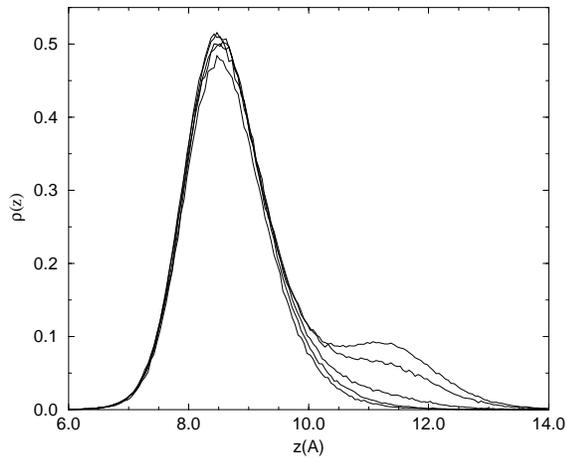}}
\caption{Density profiles near layer promotion.  The coverages are
0.2794, 0.2837, 0.2879, 0.2964, and 0.3005 atom/$\AA^2$.
}
\label{fig:promote}
\end{figure}

Promotion to the fourth layer occurs for coverages greater than 
0.2837 atom/$\AA^2$ (17 active atoms).  This coverage may be
determined from the following
observation, which is illustrated in Fig.\ \ref{fig:promote}. 
For coverages at and just below promotion, 
increasing the coverage increases
the peak height but does not appreciably change the profile's width.  
Just above this density, the peak height does not change, but an abrupt
increase in the width is observed.  With increasing coverage,
the tail of the profile just above layer promotion evolves into the
fourth layer peak.  This is also illustrated in the figure.
This value for layer promotion is in agreement with experiment.
Heat capacity measurements\cite{greywall91,greywall93} show layer promotion at
0.288 atom/$\AA^2$, while
isothermal compressibility measurements \cite{zimmerli92}
give a somewhat lower value 0.280 atom/$\AA^2$ for the promotion density.

\begin{figure}[htp]
\epsfxsize=\figwidth\centerline{\epsffile{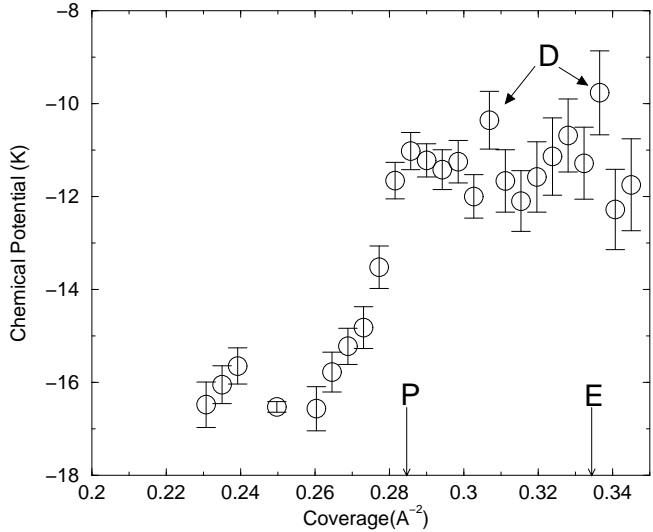}}
\caption{ Chemical potential for third and fourth layers.  The arrows
indicate the third layer promotion density (P) and the fourth layer
equilibrium coverage (E).  The densities denoted by ``D'' and the
dark arrows are discussed in the text.
}
\label{fig:chemfr}
\end{figure}

Layer promotion is also
signaled by a change in the density dependence of the chemical potential.
By differencing our calculated total energy values, we
can obtain the chemical potential $\mu$.  This is plotted in 
Fig.\ \ref{fig:chemfr}.  The values for the energy per particle
are given in Table \ref{tab:eng3fr}.
As can be seen from the figure, 
the chemical potential increases rapidly just below layer promotion
until it
becomes favorable to promote an atom to the next layer.  Above layer
promotion, the chemical potential remains 
roughly constant with increasing
density, as it should for the liquid-gas coexistence region.  This change
in the density dependence of $\mu$, together with the density profiles,
determines the promotion density.

\begin{table}
\begin{center}
\vspace{.2in}
\begin{tabular}{|c|c|c||c|c|c|} \hline
N & $\sigma$($\AA^{-2}$) & E/N (K)  & N & $\sigma$($\AA^{-2}$) & E/N (K) \\ \hline
4 & 0.2286 & -14.951(102) & 20 & 0.2964	& -14.774(18) \\
5 & 0.2329 & -15.308(56) & 21 & 0.3006	& -14.602(13) \\
6 & 0.2371 & -15.410(48) & 22 & 0.3048	& -14.479(17) \\
7 & 0.2413 & -15.426(37) & 23 & 0.3091	& -14.307(22) \\
11 & 0.2583 & -15.848(34) & 24 & 0.3133	& -14.193(18) \\
12 & 0.2625 & -15.897(24) & 25 & 0.3175	& -14.105(19) \\
13 & 0.2667  &	-15.878(24) & 26 & 0.3218 & -14.015(23) \\
14 & 0.2710  & 	-15.848(17) & 27 & 0.3260 & -13.905(21) \\
15 & 0.2752  &  -15.771(25) & 28 & 0.3302 & -13.787(19) \\
16 & 0.2794  & 	-15.624(16) & 29 & 0.3345 & -13.707(20) \\
17 & 0.2837  & 	-15.401(18) & 30 & 0.3387 & -13.573(23) \\
18 & 0.2879  & 	-15.153(14) & 31 & 0.3429 & -13.528(17) \\
19 & 0.2921  & 	-14.941(13) & 32 & 0.3472 & -13.478(27) \\ \hline
\end{tabular}
\caption{Energy/particle versus coverage at 400 mK for the third layer
when the second layer is frozen.  All calculations use the 
$15.075 \AA \times 15.667 \AA$
simulation cell.  The number in parenthesis gives
the error in the last two digits.}
\label{tab:eng3fr}
\end{center}
\end{table}

Layer promotion in general is related to the chemical potential.  As the
density of a particular layer increases above the equilibrium 
coverage, so does $\mu$.
Each atom added to the system must choose whether to go
in the dense layer or sit above it in the unfilled layer.  This choice
is governed by the chemical potentials for the two layers.  Above 
the equilibrium coverage of a layer but below promotion, the $\mu$ of the
dense layer is lower than $\mu$ for the unoccupied
layer.  However, the chemical potential for the dense layer 
increases rapidly with increasing density and at some coverage,
the chemical potential for adding an atom to the dense layer
exceeds the chemical potential for adding it to the unoccupied layer.
The above discussion is illustrated for the third and fourth layers
by Fig.\ \ref{fig:chemfr}.  As can
be seen in the range 0.26 to 0.28 atom/$\AA^2$
each additional atom
causes a rapid increase in $\mu$.  Around 0.284 atom/$\AA^2$, 
$\mu_3$ of the third
layer exceeds $\mu_4$ of the fourth layer, and so the next atom 
added to the system will
be preferentially promoted to the fourth layer.

\begin{figure}[htp]
\epsfxsize=\figwidth\centerline{\epsffile{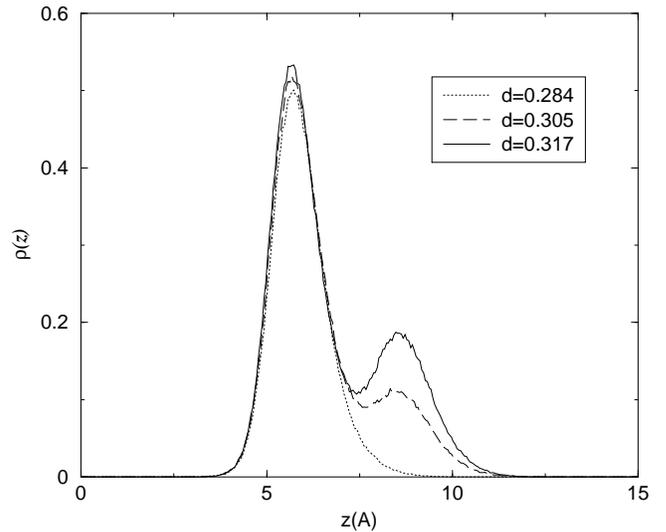}}
\caption{ Density profiles for the third and fourth layers.  Coverages
are given in $\AA^{-2}$.
}
\label{fig:demote}
\end{figure}

The beginning of particle promotion to a new layer does not signal the
end of the filling of the old layer.  As can be seen in 
Fig.\ \ref{fig:chemfr}, adding atoms to the system 
at the densities 0.306 and 0.334 atom/$\AA^2$, marked
by the letter ``D'' and the dark arrowheads, causes the chemical potential
to increase above the chemical potential at layer promotion.  Immediately
above both of these densities, $\mu$ drops back to values equal to
$\mu$ at promotion.  This signals particle demotion to the third layer.
Figure\ \ref{fig:demote} illustrates this for densities around 0.306.
Notably, each additional demoted atom increases the energy required
to further increase the layer density, so $\mu$ at promotion is only a
rough estimate of the chemical potential required to demote additional
atoms.  

\begin{figure}[htp]
\epsfxsize=\figwidth\centerline{\epsffile{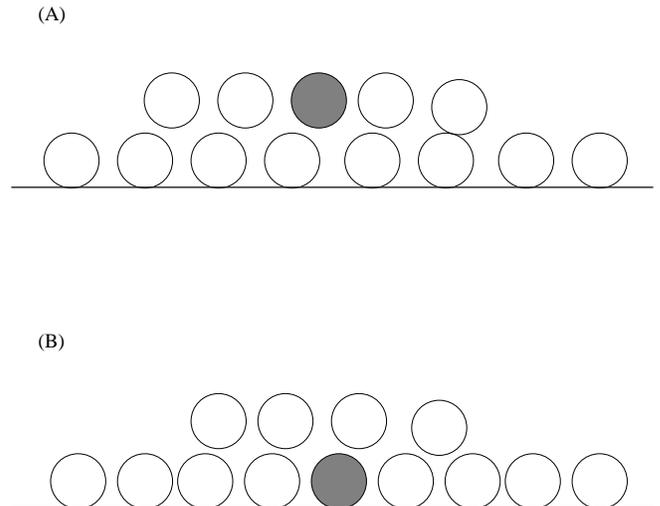}}
\caption{ An illustration of the circumstances of particle demotion.  The
shaded atom is added to the layered system.  In (A), it joins the
less dense outer layer.  In (B), it is demoted to the more dense 
underlayer.
}
\label{fig:pancake}
\end{figure}

Particle demotion may be understood as a balance between many factors.
Atoms are initially promoted above a layer because this is 
energetically favorable.  The first promoted atom loses the large
energy benefit for being close to the substrate, but gains kinetic energy
since it is free to move about on the open surface, with its wave
function no longer constrained by the other atoms.  It also
retains some of the potential energy advantage gained from having
neighboring helium atoms.  
Adding more atoms to the system will favor the formation of 
droplets.
However, once droplets have formed in the
new layer, the next atom added to the system faces a
different choice than the first promoted atom.  This is illustrated in 
Fig.\ \ref{fig:pancake}.  If it goes into the less dense outer layer,
it gains some attraction from the other atoms in this layer.  However,
it no longer gains the kinetic energy advantage that the first promoted
atom had.  On the other hand, if the added atom goes into the dense lower
layer, it regains the benefit of being closer to the strongly attractive
substrate.  Furthermore, it has the energy advantage for having more 
helium neighbors:  those in its layer, those in the layer above, and
those in the layer below (not shown).  In contrast, if the added 
atom goes into the outer layer, it has less neighbors.  It should
be noted that each demotion to the dense layer does significantly increase
the chemical potential for this layer, 
so at some point it will be more favorable
for atoms to be added to the outer layer again.


\subsubsection{Third and fourth layer phases}
We expect four principal regions in the third layer 
before fourth layer promotion.
These are a low density gas phase (which will have
a negligible density at low temperatures), a droplet region, a metastable
liquid region, and an equilibrium liquid phase. 
The droplet region consists of a liquid phase separated from a low-density
gas by an interface.  In the metastable region, the droplet phase is replaced
by a stretched uniform phase which has negative spreading pressure.  The
crossover from the droplet region to the stretched liquid phase occurs
at the spinodal point.
Direct evidence that the layer is liquid
comes from torsional oscillator measurements, which detect superfluidity
up to layer completion.  The isothermal compressibility
has been measured for the third and
higher layers\cite{zimmerli92} and exhibits a divergence that is 
associated with the spinodal point.
The equilibrium liquid coverage can be inferred from heat capacity
measurements.\cite{greywall91}  Below, we present evidence 
of each of these phases using several different observables.

\begin{figure}[htp]
\epsfxsize=\figwidth\centerline{\epsffile{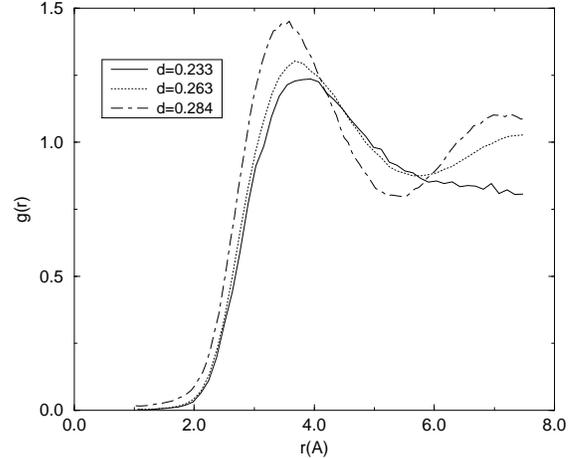}}
\caption{The radial distribution function for the third layer at
the indicated coverages, in atom/$\AA^2$.
}
\label{fig:rdf}
\end{figure}

First, we can
establish the existence of droplets and a uniform liquid phase
at different densities with the radial distribution function, g(r),
which provides a direct probe of short and long range behavior.
Calculations for the third layer are shown in 
Fig.\ \ref{fig:rdf}.  These are plotted as functions of the magnitude
of the distance vector between pairs of atoms, projected onto the
plane of the substrate.
These calculations of the averaged $g(r)$ 
smooth out possible anisotropies induced by the 
corrugations of the underlying solid helium layer.
The g(r) for the three coverages shown in Fig.\ \ref{fig:rdf} 
are representative of 
the droplet region, the equilibrium liquid, and the liquid near 
layer promotion.  At the
lowest coverage, 0.233 atom/$\AA^2$ (5 active atoms),
the radial distribution function drops below unity at large distances, 
as would be expected for a droplet phase.  The actual dimensions of the
droplet for a given density depend on the size of the simulation cell.
For the intermediate coverage, 
0.2624 atom/$\AA^2$ (12 active atoms), the
first peak has changed only slightly, but the long range behavior is
noticeably different, rising again past unity instead of dropping continuously.
At the highest coverage, 0.284 atom/$\AA^2$, the system shows evidence of
increased correlation, but the long range behavior cannot be determined
due to the small size of the simulation cell.  This largest coverage
also has a different short range behavior, showing an increased probability
that the projected distance between two atoms 
will be less than 2.0 $\AA$.  In part, this is a result of the 
thickening of the layer, as can be seen in Fig.\ \ref{fig:zdist}.  Because
the atoms can be at different heights above the substrate, the projected
distance between them can become smaller than would be
possible in strictly two-dimensional calculations.

\begin{figure}[htp]
\epsfxsize=\figwidth\centerline{\epsffile{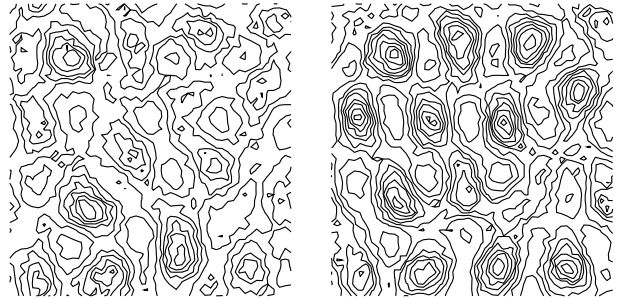}}
\caption{ Probability densities for the third layer liquid near equilibrium
(0.2625 atom/$\AA^2$, left) and just before layer 
promotion (0.2837 atom/$\AA^2$, right).
}
\label{fig:distrib}
\end{figure}

We can also gain some insight into the layer phases by examining contour
plots of the probability distribution of atoms in the plane of the
substrate.  Plots near equilibrium and layer promotion are shown in 
Fig.\ \ref{fig:distrib}.  The high density
liquid shows noticeably more localization than the equilibrium
fluid.  An increased correlation can also be seen in the radial
distribution function at high density.  This suggests that the
system may be nearing a liquid-solid coexistence phase.  

\begin{figure}[htp]
\epsfxsize=\figwidth\centerline{\epsffile{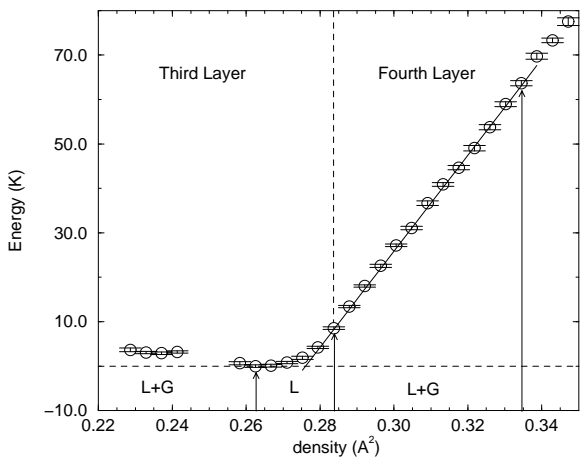}}
\caption{Shifted total energy for the third and fourth layers.  The horizontal
dashed line is the gas-liquid coexistence line for the third layer.  
The solid line is the gas-liquid coexistence region for the fourth layer.
The vertical dashed line gives the density of layer promotion.  
Gas-liquid (L+G) and uniform liquid (L) coverage ranges are indicated 
for both layers.
}
\label{fig:phase}
\end{figure}

Having established that the third layer has gas-liquid and
uniform liquid phases, we can next determine the equilibrium
coverage ranges of these phases.  This can be done by using 
the Maxwell construction.  
At low temperatures, the total energy and the total free energy are
nearly equal, so coexistence regions may be identified by applying
the Maxwell construction to the energy.  
The results for the low temperature (400 mK) temperature scans are 
shown in Fig.\ \ref{fig:phase}.  
We have verified that the energy values shown
are effectively zero temperature results by recalculating 
some values at
500 mK.  In all cases the calculations at the two temperatures agreed within 
error bars.  The values shown in the figure have been shifted
by the amount $N_{act}e_0$, where $e_0=-15.897 \pm 0.024$ is the minimum 
energy per particle, and $N_{act}$ is the number of active atoms in the
simulation.  At low temperatures,
the gas phase will have zero coverage and thus zero total energy.  We can
thus draw a coexistence line between the beginning of the third layer, 0.2117
atom/$\AA^2$, and the coverage with minimum energy per particle.
This higher coverage is the equilibrium liquid density.  We find this
coverage to be 0.2625 atom/$\AA^2$ 
($N_{act}=12$).  The best chi-squared parabolic fit around this minimum
gives 0.2645(9) atom/$\AA^2$.  
The number in parenthesis is the 
error in the last digit.  At this density the layer is completely
covered by a uniform liquid.  Below this value,
the system enters the gas-liquid coexistence region.  
The energy values in the 
coexistence region lie above the coexistence line, either because 
the liquid phase is unphysically uniform,
or else because of the appreciable cost for creating a phase 
boundary in a finite-sized system.  The
third layer equilibrium coverage, 0.0528(9) atom/$\AA^2$, is comparable
to (and slightly higher
than) the equilibrium coverage found for the second 
layer, 0.0480(6) atom/$\AA^2$.\cite{pierce98,pierce99a}
For both layers, simulated with the same size cell, the energy minimum
occurs when the system contains 12 atoms.

The equilibrium density that we determine is in good agreement with both
heat capacity and torsional oscillator measurements.  In the measurements
of Greywall, \cite{greywall91,greywall93} the low temperature
heat capacity depends linearly
on density from the beginning of the third layer to 0.260 atom/$\AA^2$.  This
linear dependence is a signal of phase coexistence.\cite{dash75}  The 
torsional oscillator
measurements\cite{reppy96} provide evidence of a similar region.  The 
temperature of the oscillator's dissipation peak, which gives a rough 
estimate for the
superfluid transition temperature, is independent of coverage from 
0.22 to 0.26 atom/$\AA^2$.  This is characteristic of a surface covered
by liquid droplets.\cite{dash78a}  Increasing the coverage in the droplet
region increases the size of the droplets, not their density, and so the
transition temperature remains constant.  

For densities above the equilibrium coverage and below layer promotion,
we have not been able to rule out the onset of solid-liquid coexistence
in the third layer.  Our procedure for identifying such 
regions in the second layer
was to look for the instabilities in the total energy that signal phase
coexistence.  This was possible because that layer solidified before
layer promotion.  In the present case, though, the energy cannot be
used to find solid-liquid coexistence because any instabilities 
that might signal solidification are 
inextricably entangled with the liquid-vapor phase coexistence that
occurs in the fourth layer at the same coverages.  Specifically,
layer promotion begins at the third layer coverage of
0.0720 atom/$\AA^2$.  The next increment in coverage that we can
simulate using the cell described in Sec.\ \ref{sec:method} is 
0.0762 atom/$\AA^2$.
In our previous simulation of the second layer
using the same sized cell,\cite{pierce98,pierce99a}
we determined that an incommensurate solid begins
to form at the second layer coverage 0.0762 atom/$\AA^2$.  Apparently,
promotion preempts solidification.
Also of note is that the density profile of the third layer 
at 0.0720 is much less peaked than the second layer at the same
layer coverage, so any third layer solid phase must have very
large zero-point motion.  One possible consequence of the third layer
entering solid-liquid coexistence is that it may fully solidify under
compression of further adsorbed layers.  As we have discussed in
the previous section, the third layer density
continues to increase even after fourth layer promotion. Layer
promotion is not a phase transition and can occur whenever the
chemical potential of the system favors it.  Thus it is possible
to have layer promotion in the middle of a phase transition.  The 
third layer solid-liquid phase transition can be completed after fourth
layer promotion as additional atoms added to the system get demoted
to the third layer.  We note finally that there is some experimental
evidence suggesting that third layer solidification may occur.
In the heat capacity results presented by Greywall, \cite{greywall93}
a small peak at about 1.8 K can be observed for coverages beginning
at 0.3100 atom/$\AA^2$, between the rounded heat capacity feature
associated with the fourth layer liquid and the sharp
peak associated with the melting of the second layer solid.  

The fourth layer also exhibits a self-bound liquid coverage.  We can
identify this in the same manner as before.
Since promotion to the fourth layer occurs above 
0.2837 atom/$\AA^2$, we can consider this coverage to correspond to a
zero density fourth layer gas.  We apply the Maxwell construction again
to the fourth layer and determine a bound liquid phase at 
0.3345 atom/$\AA^2$ (29 active atoms),
giving a fourth layer equilibrium density of 0.0508 atom/$\AA^2$.  The
solid line in the figure gives the maximum possible range for gas-liquid 
coexistence in the fourth layer.  All energy values in this region 
are on or above the coexistence line. 
This coexistence region agrees with heat capacity 
measurements,\cite{greywall93}
which exhibit linear isotherms in the fourth layer up to
0.3300 atom/$\AA^2$ at low temperatures.  

A trend can be noticed in the equilibrium density as one progresses from 
the inner to outer layers.  For the first layer of helium adsorbed on
a flat substrate, we have calculated the equilibrium density to be 
0.0450(6) atom/$\AA^2$, close to the two-dimensional value.\cite{whitlock88}
Successive increases are observed in the second and third layers, as 
the zero-point motion of the layers perpendicular to the plane of the 
substrate becomes larger.  From a two-dimensional point of view, this 
motion has the effect of softening the hard cores of the helium atoms.

\begin{table}
\begin{center}
\vspace{.2in}
\begin{tabular}{|c|c|c|} \hline
N & $\sigma$($\AA^{-2}$) & E/N (K) \\ \hline
7 & 0.2291 &    -15.199(51) \\
8 & 0.2313 &	-15.362(47) \\
12 & 0.2404 & 	-15.442(33) \\
15 & 0.2472 &	-15.689(36) \\
18 & 0.2540 &	-15.885(36) \\
24 & 0.2676 & 	-16.005(27) \\
26 & 0.2722 & 	-15.908(25) \\
28 & 0.2767 &	-15.833(30) \\ \hline
\end{tabular}
\label{tab:eng7x4}
\caption{Energy/particle versus coverage at 400 mK above the frozen
second layer using the $21.015\AA \times 20.889 \AA$
simulation cell described in the text.
The number in parenthesis gives
the error in the last two digits.}
\end{center}
\end{table}

In Fig.\ \ref{fig:phase}, we do not plot energy values for the coverages
between 0.2413 and 0.2583 atom/$\AA^2$ at 400 mK.  Calculations that we 
performed for these densities at 500 mK actually showed an energy decrease,
outside of error bars.
We attribute this to variations inherent in using frozen configurations for
the second layer:  the
two calculations required two different configurations,
and so there will be some systematic difference in the energy values.

\begin{figure}[htp]
\epsfxsize=\figwidth\centerline{\epsffile{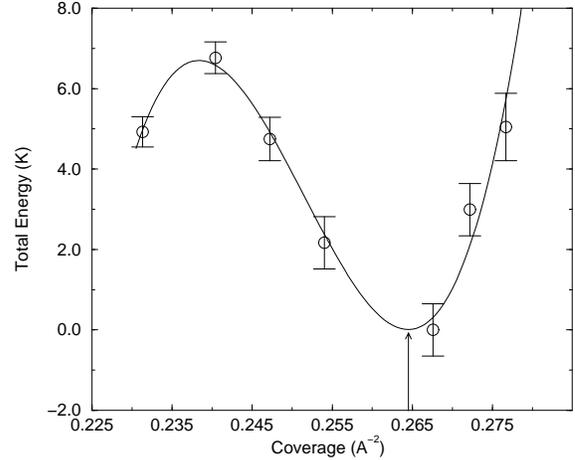}}
\caption{Total energy for the third layer using the larger
simulation cell described in the text.  
The solid curve is a fit to Eq.\ (\ref{eq:poly2}).
The arrow indicates the equilibrium liquid coverage.
}
\label{fig:shift7x4}
\end{figure}

We have repeated the third layer calculations 
at 400 mK using a larger simulation
cell ($21.105 \AA \times 20.889 \AA$).  This will allow us
to examine finite-size effects and the effects of different configurations.
We will also be able to examine the intermediate region that we 
excluded in the calculations using the smaller cell. 
The total energy values are given in Fig.\ \ref{fig:shift7x4}.  
These
energy values have been shifted by subtracting out the gas-liquid
coexistence line, as described above.  
The energy per particle values used to construct this figure
are given in Table \ref{tab:eng7x4}.  The minimum energy per 
particle is $-16.005 \pm 0.027$ K,
which occurs at 0.2676 atom/$\AA^2$ (24 atoms).  
We fit these
energy values to a polynomial in the form
\begin{eqnarray}
E/V=e_0 + B(\frac{\rho-\rho_0}{\rho_0})^2 + C(\frac{\rho-\rho_0}{\rho_0})^3,
\label{eq:poly2}
\end{eqnarray}
where $\rho$ is the coverage.
Fitted parameters are given in Table \ref{tab:fit7x4}.  
Not all digits are
significant.  Notable from the table is that the equilibrium liquid
coverage occurs at the same density as in previous calculation
with the smaller cell,
despite the fact that there is a small
discrepancy in the energy values at similar coverages between the cells.
In our previous calculations on the second layer, we observed that 
finite-size effects on the energy were negligible.  Thus we believe that
the difference in the minimum energy values obtained with the two cells
is attributable to the 
particular frozen second layer configurations that we used.
We also observed that a calculation using the larger cell but at 500 mK
and 0.2540 atom/$\AA^2$ (18 active atoms) showed an energy increase
over the value obtained at 400 mK.

\begin{figure}[htp]
\epsfxsize=\figwidth\centerline{\epsffile{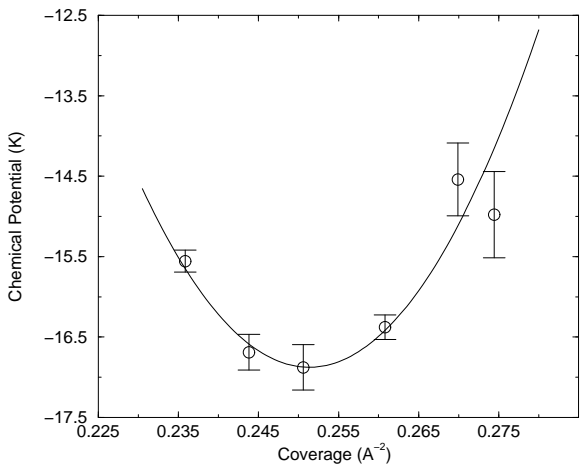}}
\caption{Chemical potential calculated from the total energy values
and the fit shown in Fig.\ \ref{fig:shift7x4}.
}
\label{fig:chemical}
\end{figure}

By differencing the (unshifted) total energy values of the larger
cell, 
we can obtain the chemical potential $\mu$.  Values
obtained from both the fit and the actual energy values 
are shown in Fig.\ \ref{fig:chemical}.  The size of the chemical
potential near the minimum ($-16.88 K \pm 0.28$) is in general agreement
with the value range of -9 to -16 K obtained for the third layer 
by Clements, {\em et al}.\cite{clements93b}  As noted
by those authors, their values are sensitive to the attractiveness of 
their model substrate.  

\begin{table}
\begin{center}
\begin{tabular}{|l r|} 
Parameter	   	& Value \\ \hline
e$_0$(K)           	& $0.0097 \pm 0.51$ \\
$\rho_0$($\AA^{-2}$)  	& $0.2645 \pm 0.0005$\\
B(K)               	& $2057.3 \pm 237.0$ \\
C(K)		   	& $13885.7 \pm 1759.0$ \\
$\chi^2/\nu$       	& 1.09 \\ 
\end{tabular}
\caption{ Fitted parameters for the polynomial fit to
the shifted total energy values.}
\label{tab:fit7x4}
\end{center}
\end{table}

We can determine the spinodal point by taking the second derivative of 
Eq.\ (\ref{eq:poly2}).  The isothermal compressibility is given
by $\kappa_T = \rho^{-2}(\partial \mu/\partial \rho)_T^{-1}$, 
where $\rho$ is the
coverage.  This diverges when the derivative of the chemical potential
is zero.  Below this coverage, $\partial \mu/\partial \rho$ is negative and
the speed of sound becomes imaginary.  This is the spinodal point.  
From the fit, we determine that this occurs at 0.2518 atom/$\AA^2$,
a third layer coverage of 0.0386 atom/$\AA^2$.  This 
is in agreement
with the HNC-EL calculations for the third layer.\cite{clements93b}
Experimentally, the spinodal point can be determined from the divergence
of the isothermal compressibility.  In the third layer,
this occurs near the third layer coverage of 0.03 atom/$\AA^2$ 
at 1.2 K.\cite{zimmerli92} 

\begin{figure}[htp]
\epsfxsize=\figwidth\centerline{\epsffile{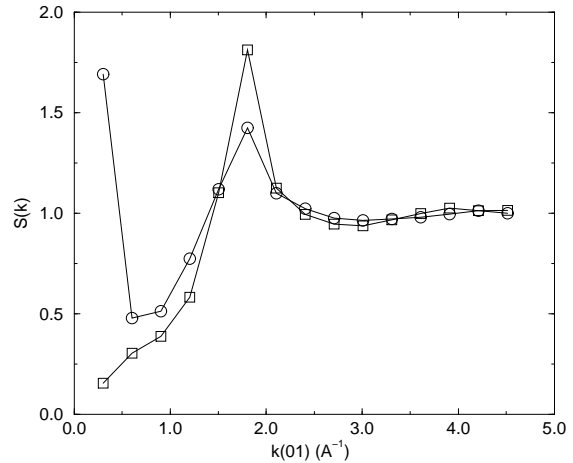}}
\caption{Static structure factor at 0.2472 (circles) and 
0.2676 atom/$\AA^2$ (squares) at T=400 mK.
}
\label{fig:ssf7x4}
\end{figure}

The droplet region and the uniform coverage region for this
simulation cell can be directly identified by
the short wavelength behavior of the static structure function, S(k).  
Figure \ref{fig:ssf7x4} shows S(k) for the coverages 0.2472 (below the
spinodal point) and 0.2676 atom/$\AA^2$ (near equilibrium).  The upward
swing of $S(k)$ for the lower density for small values of $k$ indicates
the presence of a droplet.  Near the equilibrium density, the
droplet has entirely covered the substrate and so 
$S(k \rightarrow0) \rightarrow 0$.

\begin{figure}[htp]
\epsfxsize=\figwidth\centerline{\epsffile{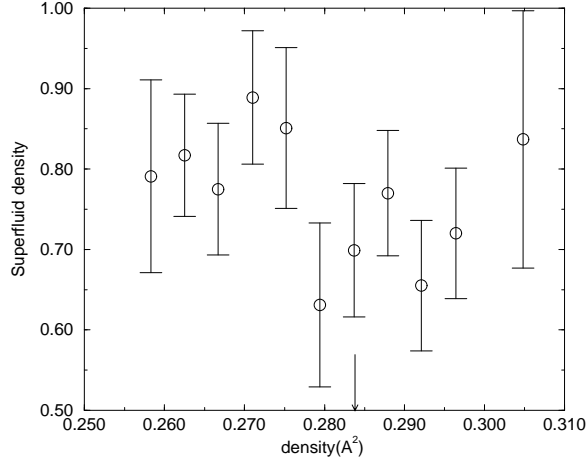}}
\caption{ Superfluid density values at 400 mK.  The fraction is relative to 
the number of active atoms.  The arrow indicates the density at which 
fourth layer promotion begins.
}
\label{fig:super}
\end{figure}

At the temperature of our simulation, superfluidity can be observed in
the third and fourth layers.  Superfluid density values, ranging from
intermediate third layer to low fourth layer coverages, are shown in 
Fig.\ \ref{fig:super}.  The values shown are the ratios of the
superfluid density to the third layer coverage.
A slight suppression in superfluidity occurs for 
coverages just below and just above the promotion density to the fourth layer,
although the noise of our calculations obscures this effect.
Suppression of superfluidity 
has been observed at similar temperatures and coverage values 
in both the torsional oscillator measurements of Crowell and
Reppy\cite{reppy93,reppy96} and a simulation of helium on 
hydrogen.\cite{wagner94b} 
Suppression just after fourth layer promotion is a consequence of gas-liquid
coexistence.  In the low coverage region, the fourth layer consists of droplets
that lack the connectivity to exhibit superfluidity until they have
percolated across the surface.\cite{dash78a,campbell97}
The suppression of superfluidity before promotion to the fourth layer
can be observed
more clearly when we allow the second layer to respond to the
third layer film, as described in the next section.

\subsection{Results with active second layer}
\subsubsection{Layer promotion and demotion}
In the previous calculations we assumed that the solid second layer could
be treated as inert; that is, the response of the second layer 
to the third was ignored.  It is known experimentally, however, 
that the second
layer solid continues to be compressed after third layer promotion.
Evidence of this can be found in the heat capacity peak
associated with second layer melting.\cite{greywall93} 
This peak continues to increase in temperature for the entire 
range of the third
layer, indicating an increase in the second layer density.  Neutron
scattering\cite{carneiro81,lauter91a} also detects
compression of the second layer by the third, with an abrupt restructuring
at intermediate third layer coverages.
Second layer compression is 
not unexpected, since the compression of the first layer by the growth
of the second is well established.\cite{bretz78} 
To allow for the effects of second layer compression, as well as
the response of this layer to the growth of the third, we have
performed calculations that include both second and third layer
atoms in the Monte Carlo sampling.  

\begin{figure}[htp]
\epsfxsize=\figwidth\centerline{\epsffile{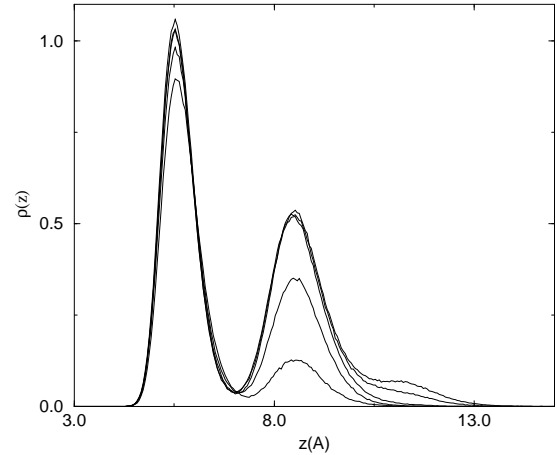}}
\caption{Density profiles for the active second and third
layers as a function of height above the substrate.   The coverage values
are 0.2296, 0.2625, 0.2879, 0.2964, and 0.3045 atom/$\AA^2$.
}
\label{fig:zdistnew}
\end{figure}

Density profiles illustrate the compression of the second layer in
our simulation.  These
are shown in Fig.\ \ref{fig:zdistnew} for selected coverages.  The left 
and right peaks in the figure
are for the second and third layers, respectively.  As can be seen, the 
height of the second layer peak grows between the coverage 0.2296 and
0.2879 atom/$\AA^2$.  The peak 
height then remains constant up to and including the
coverage 0.2964, where promotion to the fourth layer is visible.  
Finally, the second layer peak increases again at the highest coverage 
examined, 0.3049 atom/$\AA^2$.

By integrating the profiles up to the minimum between
the two peaks (at approximately 7.5 $\AA$), we obtain
the second layer coverages of 0.0845, 0.0885, 0.0894, 0.0890,
and 0.0929 atom/$\AA^2$ for the 
coverages shown in the figure.  Thus at the lowest 
third layer densities, no demotion occurs.
Beginning at intermediate coverages and up to layer promotion,
a single atom is demoted.  At low fourth layer coverages (the highest 
coverage examined) two atoms are demoted.  
As a consequence of demotion, promotion to the fourth layer is pushed
to a higher overall coverage.  As is illustrated in Fig.\ \ref{fig:zdistnew},
we do not observe promotion to the fourth layer until the density has exceeded 
0.2879 atom/$\AA^2$, compared to 0.2837 atom/$\AA^2$ when the second layer
was frozen.  This value will be further established below by examining
the density dependence of the chemical potential.
This value is still in agreement with the heat capacity\cite{greywall93}
and isothermal compressibility\cite{zimmerli92,reppy96} measurements.
We note that Zimanyi, {\em et al}\cite{zimanyi94} have proposed a slightly
higher completion coverage of 0.293 atom/$\AA^2$.

\begin{figure}[htp]
\epsfxsize=\figwidth\centerline{\epsffile{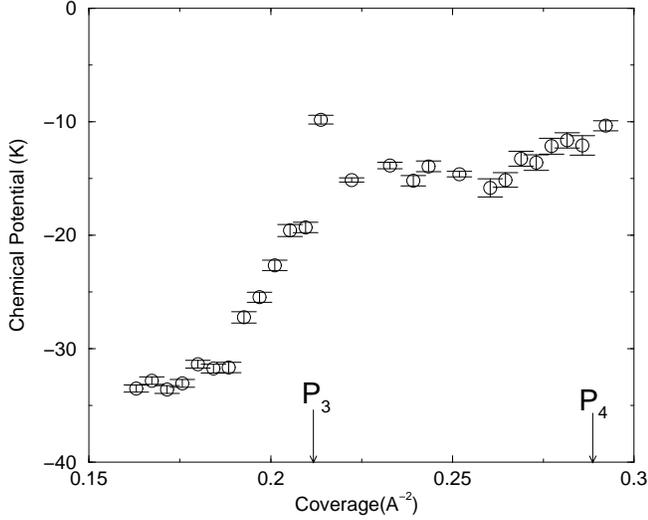}}
\caption{Chemical potential for the second and third layers.  Third
layer and fourth layer promotion densities are indicated by the
arrows.
}
\label{fig:chem5x3}
\end{figure}

The layer promotion illustrated above is accompanied by a discontinuity in the 
chemical potential.  We can obtain this quantity by differencing our 
calculated total energy values, given in Table \ref{tab:engnew}, and 
plot the results in Fig.\ \ref{fig:chem5x3}. 
Included in this figure are
results obtained for the second layer from a previous 
calculation.\cite{pierce99a}  As can be seen, there is a distinctive
change in the density dependence of the chemical potential $\mu$
around the promotion density.  Below the 
promotion density, the energy changes very rapidly with increasing
coverage.  Near promotion, an added atom has the choice of going
to the unoccupied third layer or the dense second layer, and will 
choose the layer that is energetically favorable.  When the
promotion coverage is reached,
the chemical potential for adding the atom to the second layer
exceeds that for adding it to the third layer, and so the atom
is added to the unoccupied layer.
It can be seen from the figure that the change in $\mu$ associated with the
promotion is quite large. In this case, the second layer is solid
and relatively
dense, so we expect an energy gap to be associated with promotion.
As can be seen in the density profiles (see Fig.\ \ref{fig:zdistnew}) for
this layer, there is a range of heights above the second layer that
is forbidden for the third layer atom.  This is in contrast to
what happens in the case of fourth layer promotion (see 
Fig.\ \ref{fig:promote}), in which there is significantly more 
overlap between the third and fourth layers.
Each additional atom added to the system will also
go the outer layer, but the rate of the energy change will decrease, 
since these atoms are attracted to each other and form a droplet.
Thus we see that the chemical potential just after layer promotion decreases.

\begin{table}
\begin{center}
\vspace{.2in}
\begin{tabular}{|c|c|c|} \hline
N & $\sigma$($\AA^{-2}$) & E/N (K) \\ \hline
19 &	0.2075	& -30.687(19) \\
20 &	0.2117	& -30.111(13) \\
21 & 	0.2159	& -29.257(39) \\
24 & 	0.2286	& -27.391(20) \\
26 & 	0.2371	& -26.354(12) \\
27 & 	0.2413	& -25.937(12) \\
28 & 	0.2456	& -25.516(13) \\
31 &	0.2583	& -24.462(21) \\
32 & 	0.2625	& -24.188(15) \\
33 & 	0.2667	& -23.910(13) \\
34 & 	0.2710	& -23.603(15) \\
35 & 	0.2752	& -23.314(13) \\
36 & 	0.2794	& -23.001(14) \\
37 & 	0.2837	& -22.699(12) \\
38 & 	0.2879	& -22.417(19) \\
40 &	0.2964	& -21.816(13) \\ \hline
\end{tabular}
\label{tab:engnew}
\caption{Energy/particle versus coverage at 400 mK with positions of
both second and third layer atoms sampled.  Calculations used the $5 \times 3$ 
simulation cell.  The number in parenthesis gives
the error in the last two digits.}
\end{center}
\end{table}

\subsubsection{Layer phases}
\begin{figure}[htp]
\epsfxsize=\figwidth\centerline{\epsffile{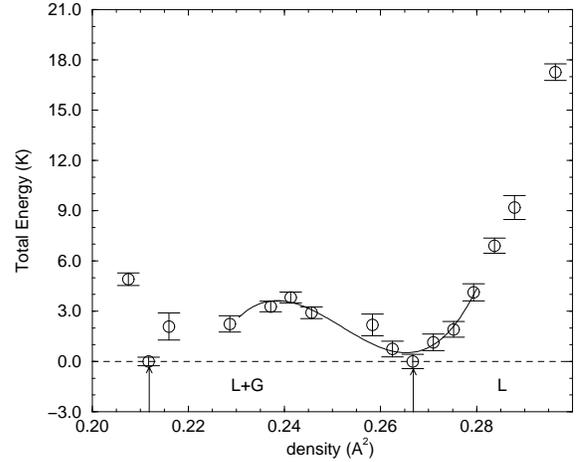}}
\caption{Total energy at 400 mK for the three layer system when second layer
atoms are included in the sampling.  The dashed line is the gas-liquid
coexistence line for the third layer.  The arrows indicate the beginning 
of promotion to the third layer
and the equilibrium third liquid coverage.  Coverage ranges
for the gas-liquid (G+L) and uniform liquid (L) phase in the third layer
are shown.  The solid line is the fit to Eq.\ (\ref{eq:poly2}).
}
\label{fig:phasenew}
\end{figure}

We can determine phase boundaries for the third layer by again 
using the Maxwell construction.  Figure \ref{fig:phasenew} gives the 
total energy values for the three layer system, with both second and
third layers active.  The energy values per particle are given 
in Table \ref{tab:engnew}.
As before, a coexistence line can be drawn between
the beginning of the third layer, 0.2117 atom/$\AA^2$ and the equilibrium
liquid coverage.  The upper endpoint of the coexistence region is 
0.2667 atom/$\AA^2$.  This is the
equilibrium coverage for the third layer liquid.  The intermediate
energy values are in gas-liquid coexistence.  Following our usual procedure,
we have subtracted out the gas-liquid coexistence line from all the energy 
values, using the values $-30.111 \pm 0.019$ K and
$-23.910 \pm 0.013$ K for the beginning (0.2117 atom/$\AA^2$, 20 atoms)
and end (0.2667 atom/$\AA^2$, 33 atoms) of the coexistence region.

As with the results found using the 
frozen second layer, we can fit the energy values to a polynomial
in the form of Eq.\ (\ref{eq:poly2}).  This is the solid curve
 in Fig.\ \ref{fig:phasenew}.
Energy values for the very low and high coverages of the layer were not
included in the fit.  The equilibrium coverage determined is
$0.2653 \pm 0.0005$ atom/$\AA^2$, and the spinodal point 
is found to be near the third layer coverage
of 0.040 atom/$\AA^2$.

\begin{figure}[htp]
\epsfxsize=\figwidth\centerline{\epsffile{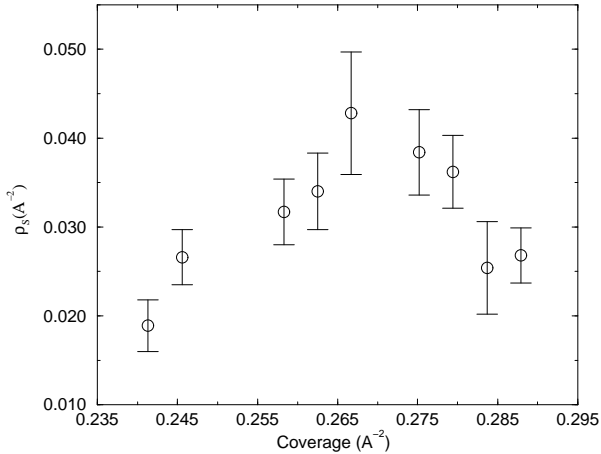}}
\caption{Superfluid coverage for the third layer at 400 mK with
both second and third layer atoms sampled.
}
\label{fig:sfnew}
\end{figure}

Suppression of superfluidity before promotion to the fourth layer is
even more prominent in these calculations.  Figure \ref{fig:sfnew} shows
the superfluid coverage (instead of the superfluid fraction) versus the 
coverage for the third layer.  For superfluid films 
adsorbed on heterogeneous surfaces (such as Vycor), the
superfluid coverage increases continuously with increasing 
coverage.\cite{reppy80}  This does not occur in the atomically thin
fluid layer we are simulating.  Superfluidity instead
increases with increasing coverage until the equilibrium density is reached, 
after which there is a plateau and then a drop.  The drop in $\rho_s$
is seen experimentally as well, for coverages greater than 0.280.

The reason for this suppression before layer promotion is not clear,
but three related factors can contribute.  First, increasing the
density in the liquid layer produces excitations that should disrupt
the superfluid state.  This energy increase can be seen in 
Fig.\ \ref{fig:phase} and Fig.\ \ref{fig:phasenew}.  
Second, increased correlation between the
atoms will suppress the permutations required for superfluidity.
We observe this increase in the radial distribution function, 
Fig.\ \ref{fig:rdf}, and in the probability contours.  The calculated
probability distributions also suggests the third mechanism, namely 
that the layer is partially solidifying before layer promotion.  As
discussed above, we have not found clear evidence that this occurs.

\section{SUMMARY}
\label{sec:summary}

This paper has presented path-integral Monte Carlo
calculations for the third and fourth
layers of helium on graphite
using two different treatments of the underlying solid layers.
In the first approach, we treated the first and second layers as
inert.  The second layer solid was initially equilibrated, and then
atoms were frozen in some particular configuration.
This simplification neglects the response of the second layer to
growth of the third layer, as well 
effects such as particle demotion, but allows us to simulate larger systems.
Using this approach, we have produced as many as three additional 
layers on top of the solid first and second layers.  The third and fourth
layers are found to possess self-bound liquid phases with layer densities 
of 0.0528 and 0.0508 atoms/$\AA^2$, respectively.  These densities
are in agreement with values that may be inferred from heat capacity and
torsional oscillator measurements.  In the third layer,
below the equilibrium density, we determine the spinodal point to be 
0.0386 atoms/$\AA^2$.  This coverage separates the unstable region, in
which droplets form on the surface, from the metastable ``stretched''
phase just before equilibrium.  Static structure calculations confirm
that droplets occur at low densities.  The chemical potential was
also calculated, and third layer promotion was found to coincide with
a discontinuity in the density dependence of this quantity.
The superfluid density has been
calculated for coverages from the third layer equilibrium to low
fourth layer coverages.  Promotion to the fourth layer is observed at
the (total) coverage 0.2837 atoms/$\AA^2$.  The continued growth of
the third layer is also observed after fourth layer promotion.
The question of partial
solidification in this layer remains open for further investigation.
In particular, it is possible that particle promotion preempts the
formation of a third layer solid, but particle demotions at higher
coverages may increase the density of this layer to the point of
solidification.  Experimentally, there is some suggestive evidence for 
third layer solidification through compression by higher layers.  
We find that the
heat capacity feature that emerges above 0.3100 atom/$\AA^2$\cite{greywall93}
might be related to the melting of this solid.  Solidification also provides
a natural explanation for the plateaus seen in torsional oscillator 
measurements of the superfluid density of the higher layers.

The effects of including the second layer atoms
in the Monte Carlo updating procedure
have also been examined.  These calculations incorporate 
the effects of zero-point motion in the second layer solid and allow this
layer to respond to the third and higher layers.  We observe particle
demotion from the third to the second layer near equilibrium and again 
near the fourth layer promotion density.  The overall coverage at 
which the third layer liquid phase forms increases slightly from 0.2645
to 0.2653 atoms/$\AA^2$
when the second layer is active.  The completion density
is also slightly higher, occuring at 0.2879 atoms/$\AA^2$.  A very
notable effect produced by the active second layer was the
reduction in the third layer superfluid density at high coverages.
This effect has also been seen experimentally.

\end{document}